\documentclass[final,1p,times]{elsarticle}

\usepackage{graphicx} % do not change
\usepackage{amssymb} % do not change
\usepackage{amsthm} % do not change
\usepackage{lineno}

\begin{document}

\title{Elliptic Flow at Large Viscosity}

\author{Volker Koch}

\address{Nuclear Science Division, Lawrence Berkeley National Laboratory,
Berkeley, CA 94720 USA}
\begin{abstract}
In this contribution we present an alternative scenario for the large
elliptic flow observed in relativistic heavy ion collisions. Motivated
by recent results from Lattice QCD on flavor off-diagonal susceptibilities
we argue that the matter right above $T_{c}$ can be described by
single-particle dynamics in a repulsive single-particle potential,
which in turn gives rise to elliptic flow. These ideas can be tested
experimentally by measuring elliptic flow of heavy quarks, preferably
via the measurement of $J/\Psi$ elliptic flow. 
\end{abstract}
\maketitle
%\linenumbers 

\section{Introduction}

One of the big surprises of the RHIC experimental program has been
the large elliptic flow \cite{Ackermann:2000tr,Voloshin:2008dg},
which, contrary to SPS energies, agreed more or less with predictions
from ideal hydrodynamics \cite{Teaney:2000cw,Kolb:2003dz}. This
surprising agreement let to the conjecture that the matter at RHIC
is a strongly coupled, nearly perfect fluid, with very small shear
viscosity. Indeed using the AdS-CFT correspondence, is was shown that
a large class of strongly coupled theories seem to have a universal
minimal shear viscosity of $\eta/s=1/4\pi$ \cite{Son:2004iv}.
Meanwhile, more refined calculations
based on relativistic viscous hydrodynamic \cite{Luzum,Song} seem to indicate
that a
finite but small value for the shear viscosity is required in order
to reproduce the $p_{t}$ dependence of the measured transverse flow.
On the theoretical side very little is known about the shear viscosity
of high temperature QCD. Perturbative calculations lead to a considerably
larger value than the conjectured lower bound. Extracting a value
for the shear viscosity from Lattice QCD (LQCD), on the other hand,
requires analytic continuation to real time, thus leading to substantial
uncertainties \cite{meyer}. It was also found that the elliptic flow of the
observed
hadrons scales with the number of quarks \cite{Voloshin:2008dg}
as predicted by a quark coalescence picture of
hadronization \cite{Voloshin:2002wa}.
This in turn implies that the scaled $v_{2}$ may be interpreted as
that of the quarks prior to hadronization, and we will use this interpretation
in the following where we always refer to the elliptic anisotropy of
quarks.

In this contribution we want to entertain an entirely different view
and interpretation of the observed elliptic flow. First, we note that
Lattice QCD results \cite{Cheng:2008zh} suggest a
quasi-particle picture, at least for the quarks. Both flavor-off-diagonal
susceptibilities \cite{Koch:2005vg} and higher order baryon number
susceptibilities \cite{Ejiri:2005wq} are consistent with vanishing
correlations for temperatures right above the transition, $T\gtrsim1.2T_{c}$.
Estimating the strength of correlations in the gluon sector is not
so straightforward, due to the lack of any additional quantum numbers,
such as flavor, which one can use to study correlations. Therefore,
let us conjecture, that gluons behave like quasi-particles as well.
In this case we have a single-particle description of the QGP right
above $T_{c}$. 

Next we need to address the equation of state above $T_{c}$, where LQCD
finds the pressure to be considerably $(\approx15\%)$ below that of
a free gas of massless quarks and gluons. Since LQCD calculations
are carried out in the grand-canonical ensemble, i.e. at fixed chemical
potential rather than particle number, a reduction of the pressure
in a single-particle picture implies a \emph{repulsive}, density dependent
single-particle potential, i.e, 
\begin{equation}
p\sim\int d^{3}p\,\exp\left[-\frac{E_{0}+U}{T}\right]<\int
d^{3}p\,\exp\left[-\frac{E_{0}}{T}\right]\sim p_{0}
\end{equation}
for $U>0$. Here $p_{0}$ denotes the pressure of a free, non-interacting
gas of partons. 

The presence of a \emph{repulsive} single-particle potential has interesting
consequences, especially for the elliptic anisotropy, $v_{2}.$ Given
the almond shaped initial distribution of matter in the transverse
plane in a semi-central heavy-ion collision, the (negative) gradient
of the potential, and thus the force, is larger in the in-plane 
than in the out-of-plane direction. As a consequence the momentum
kick due to the repulsive single-particle potential is larger in plane
than out of plane, resulting in a deformation of the momentum distribution
in qualitative agreement with the observed elliptic anisotropy, $v_{2}(p_{t})$
(see also \cite{cassing}).
We note that this effect does \emph{not} require a short mean free
path or, equivalently, a small viscosity. 

Besides the positive $v_{2}$ the single-particle dynamics
leads to two additional, qualitative predictions. First, the elliptic
anisotropy should vanish for large transverse momenta, since the 
additional momentum kick due to the potential becomes negligible.
Thus, contrary to ideal hydrodynamics we predict a maximum of the transverse
momentum dependent elliptic anisotropy, $v_{2}(p_{t})$, which is observed
in experiment. At large $p_t$, of course, $v_{2}$ will be dominated by
the attenuation of fast partons in the matter, and it needs to be
determined at what momentum this transition will take place \cite{Liao:2009zg}.
Second, since for a given temperature the momentum distribution for
heavy quarks is titled towards higher momenta, $T\sim\frac{p^{2}}{m}$,
the resulting elliptic anisotropy for heavy quarks should be considerably
smaller than that for light quarks.

\section{Schematic Model}

In order to study the qualitative features of the proposed single-particle
dynamics in a transparent fashion let us start with a simple
schematic model. If we ignore collisions among the partons, the dynamics
of the phasespace distribution follows as Vlasov equation. To allow
for an analytical treatment of the expansion, we assume a Gaussian
distribution in configuration space and non-relativistic kinematics,
resulting in a Gaussian momentum space distribution,
\begin{equation}
f(x,y,v_{x,}v_{y},t=0)=\frac{N}{2\pi^{2}\sigma_{x}\sigma_{y}(T/m)}\exp(\frac{-x^
{2}}{\sigma_{x}^{2}})\exp(-\frac{mv_{x}^{2}}{2T})\,\exp(\frac{-y^{2}}{\sigma_{y}
^{2}})\exp(-\frac{mv_{y}^{2}}{2T}).
\end{equation}
Here, $v_{x}$, $v_{y}$ are the velocities and $\sigma_{x}$, $\sigma_{y}$
denote the widths of the distribution in the transverse $x$, $y$
direction, respectively. Assuming, for simplicity, that the single
particle potential $U$ is proportional to the density of the light
degrees of freedom, \[
U\left(\vec{x},t\right)=g\,\rho\left(\vec{x},t\right)=g\int d\vec{v}\, f(\vec{x},\vec{v},t),\]
the Vlasov equation leads to the following expression for the velocity
space distribution, $n\left(\vec{v}\right)$,
\begin{equation}
\frac{\partial}{\partial t}n(\vec{v})=\frac{g}{m}\int
d^{2}x\,\vec{\nabla}_{v}f(\vec{x},\vec{v},t)\vec{\nabla}_{x}\rho(\vec{x},
t)\label{eq:transport}
\end{equation}
which can be solved analytically under the assumption that 
the time dependence of the density follows free streaming, i.e. we
ignore the effect of the potential on the density distribution. A
fully consistent solution will require a numerical treatment, which
will be briefly discussed below. Since the heavy quarks are rare, we ignore
their contribution to the potential, and propagate them in the potential
generated by the light degrees of freedom. 

To leading order in the initial spatial eccentricity
$\epsilon\equiv\frac{\sigma_{y}^{2}-\sigma_{x}^{2}}{\sigma_{x}^{2}+\sigma_{y}^{2
}}$
we then obtain the following result for the elliptic anisotropy of
the light quarks
\begin{equation}
v_{2,light}\left(u\right)=\frac{1}{2}\epsilon\frac{U_{0}}{T}\frac{1}{u^{2}}\left
[1-\exp\left(-u^{2}\right)\left(u^{2}+1\right)\right]
\end{equation}
which only depends on the kinetic energy $u=\frac{mv^{2}}{2T}$. Here,
$U_{0}=U(\vec{r}=0,t=0)$ is the initial strength of the potential
at the center. We note that $v_{2}$ is (a) proportional to the initial
spatial eccentricity, (b) proportional to the initial transverse density
(via $U_{0}$) and (c) a function of the kinetic energy only. This
is precisely what is seen in experiment. 

\begin{figure}[ht]
\centering
\includegraphics[width=0.4\textwidth]{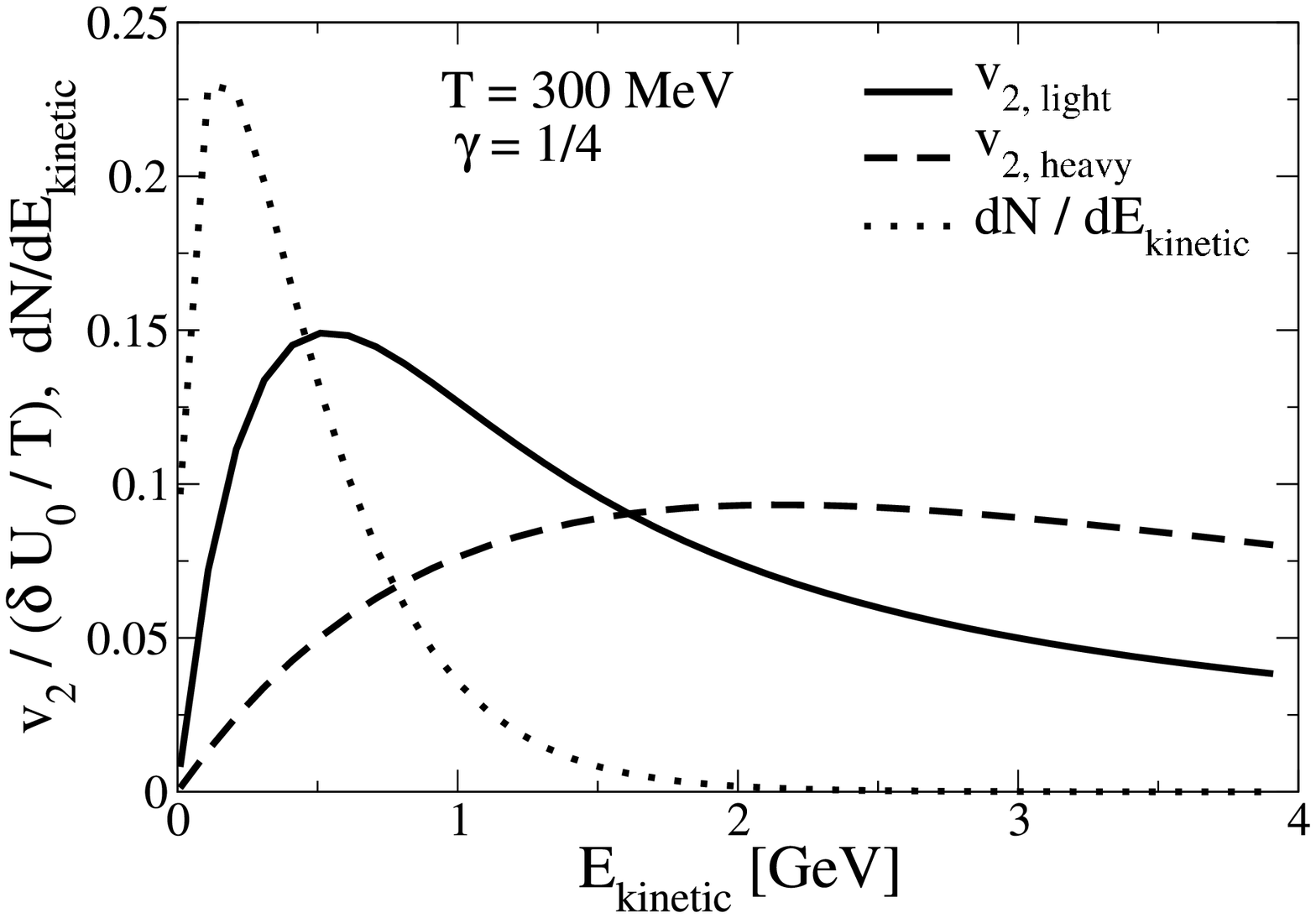}
\hspace{.5cm}
\includegraphics[width=0.4\textwidth]{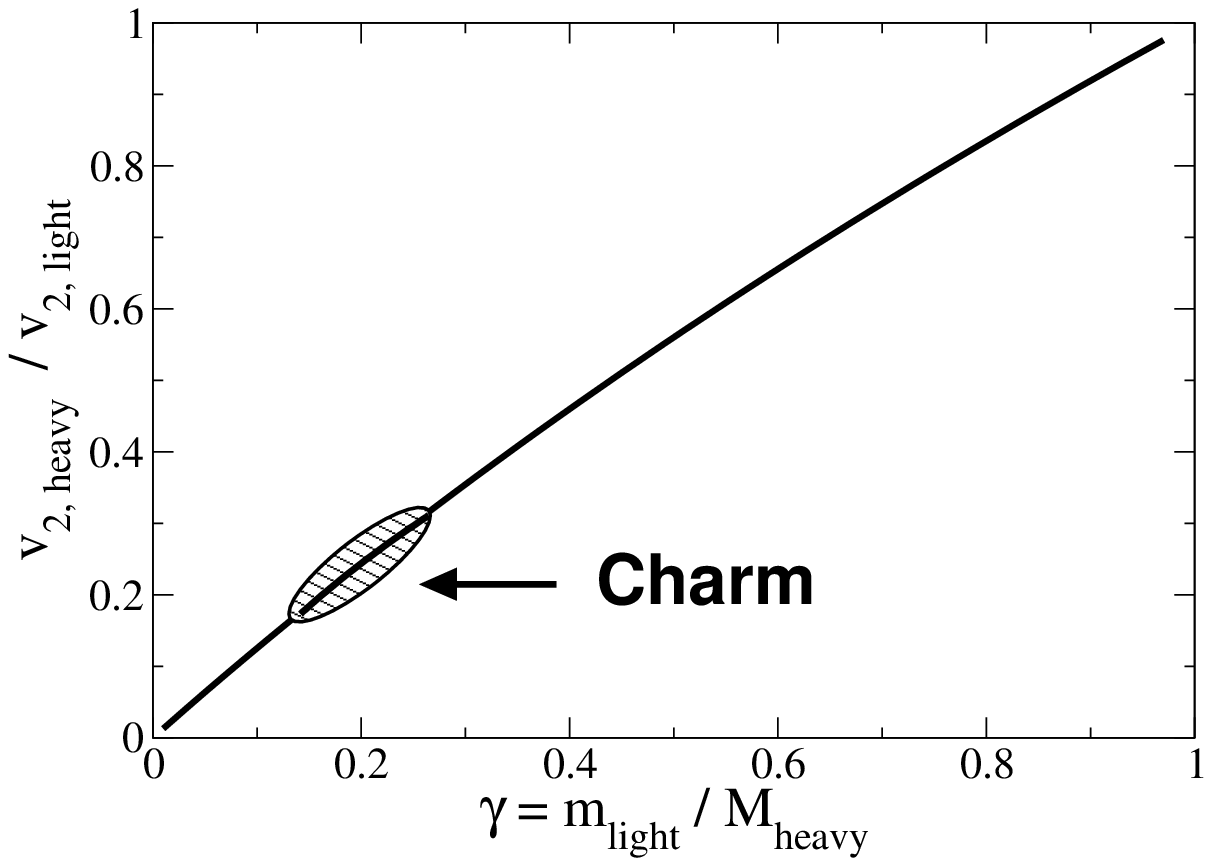}
\caption{Left panel: Schematic model results for the elliptic anisotropy as
a function of the kinetic energy, $v_{2}(E_{kinetic})$ for light
quarks (full line) and heavy quarks (dashed line). The dotted line represents
the kinetic energy spectrum for both heavy and light quarks. Right
panel: Ratio of integrated $v_{2}$ for heavy over light quarks as
a function of the ratio of heavy and light quark mass.}
\label{fig:supress} 
\end{figure}

At a given velocity the heavy quark elliptic anisotropy, $v_{2,heavy}$,
is related to that of the light quarks by $v_{2,heavy}(v)=\frac{1+\gamma}{2}\,
v_{2.light}(v)$.
Both, $v_{2,light}$ and $v_{2,heavy}$ are plotted as a function of the kinetic
energy in the left panel of Fig.\ref{fig:supress} for an
initial temperature of $T=300\,{\rm MeV}$. As expected $v_{2}$ exhibits
a maximum, the position of which depends on the choice of initial
temperature and is shifted to larger values of the kinetic energy
for  heavy quarks. As a result the momentum averaged, or integrated
anisotropy, $\bar{v}_{2}$, for the heavy quarks is much smaller. This
is shown in left panel of Fig.\ref{fig:supress}, where we plot the
ratio of heavy over light quark $v_{2}$ as a function of the ratio
of the quark masses, $\gamma=m_{light}/M_{heavy}$ . Thus, a unique
prediction of the single-particle dynamics is that the integrate $v_{2}$
for heavy quarks, specifically the $J/\Psi$ should be considerably
smaller than that for the pions, contrary to hydrodynamics where they should be
about equal. \vspace{.2cm}

\section{More realistic (transport) model}
\begin{figure}
\begin{centering}
\includegraphics[width=0.4\textwidth]{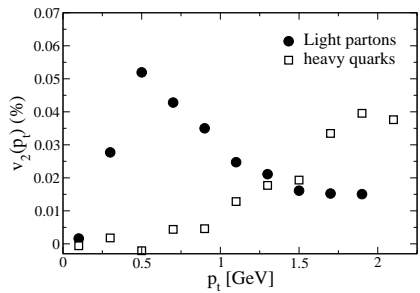}
\par\end{centering}

\caption{Result for elliptic anisotropy from transport calculation without
parton
scattering.}
\label{Flo:transport}
\end{figure}

A realistic treatment of the proposed single-particle dynamics requires
relativistic kinematics as well as a fully consistent
treatment of the Vlasov (Boltzmann) equation. In order to ensure energy
momentum conservation it is best to start from an effective energy
functional which is tuned to reproduce the Lattice equation of
state and from which one derives the single-particle Vlasov equation.
The results from such an exercise (for details see \cite{Koch_soon}) is shown in
Fig. \ref{Flo:transport}. Again we see the same features as in the
schematic model, and we find that the intergrated $v_{2}$ for heavy
quarks is considerably smaller than that of light quarks.

\section{Conclusions}

In this contribution we have shown that the qualitative features of
the observed elliptic anisotropy can be obtained from single-particle
dynamics motivated by recent Lattice QCD results. In
this description
there is no need for a very short mean free path or, equivalently,
small viscosities. This model can be tested in experiment by measuring
the elliptic anisotropy of heavy quarkonium, which is predicted to
be considerably smaller than that of pions, in contradistinction
to the predictions from hydrodynamics.

\section*{Acknowledgments}

This work is supported by the Director, Office of Energy Research,
Office of High Energy and Nuclear Physics, Divisions of Nuclear Physics,
of the U.S. Department of Energy under Contract No. DE-AC02-05CH11231.

\end{document}